\documentclass[10pt]{article}
\usepackage{amsmath, amsfonts, amssymb}
\usepackage{graphics}
\renewcommand{\vec}[1]{ {\mathbf #1} }
\begin{document}

\title{Variations of decay rates of radio-active elements
 and their connections with global anisotropy of physical space}
\author{Yu.A.Baurov$^1$, I.F.Malov$^2$\\
$^1$ Central Research Institute of Machine Building,\\ 141070,
Pionerskaya, 4, Korolyov, Moscow region 
\\
$^2$ Pushchino Radio Astronomy Observatory, P.N.Lebedev Physical
Institute,\\ Russian Academy  of Sciences, Pushchino, Moscow
Region, 142290, Russia}

\date{}
\maketitle
\begin{abstract}
The analysis of correlations between fluctuations of $\alpha$- and
$\beta$-decay rates for different radio-active elements and
Earth-Sun distance is carried out. These fluctuations exceed
significantly errors of measurements in many cases. They have the
periodical character and reveal definite spatial directions. We
suggest that the observed fluctuations are caused by the unique
physical reason connected with the global anisotropy of physical
space and by the new force.
\end{abstract}

\section{Introduction}

It is usually thought [1] that the $\beta$-decay and
$\alpha$-decay of radioactive elements are pure random processes
and their rates of decay practically are not prone to any external
influences.

For example, the intensity of $\beta$-decay can be only
insignificantly acted by high pressure. It is shown in Ref. [2]
that at a pressure of $10^5$ atm the $\beta$-decay rate for
metallic technetium is only $0.025\%$ more than at the normal
pressure. It has been experimentally well established a minor $L$-
and $K$-shells capture dependence of $\beta$-decay processes. For
example, if $^7Be$ enters into the composition of a metal, the
orbital capture goes somewhat slower than in the case when
beryllium is a constituent of oxide. The half-periods of decay due
to orbital capture differ by $0.015\%$ [2]. In a nearly 160 days
experiment with $^7Be$ introduced into a fulleren ($C_{60}$), a
difference up to $0.83\%$ in the decay half-periods between $^7Be$
in $C_{60}$ and $^7Be$ in metal was detected [3].

In Ref. [4], a non-exponential radioactive decay of nuclei-isomers
of $^{125m}Te$ during uninterrupted measurements of
$\gamma$-radiation in the course of a year, was also observed.

In recent experimental works [5-10], there were measured changes
in the $\beta$-decay and $\alpha$
- decay rates of various
radioactive elements that are considerably greater than
experimental errors. As a rule, such changes were of periodic
character [5, 6] and corresponded to certain fixed spatial
directions [7-10].

Changes in the $\beta$-decay rate are of considerable importance
in the cosmochronology [11] since the time elapsed from the start
of the Universe expansion is estimated in standard cosmology
models on the base of radioactive decay study assuming the
present-day $\beta$-decay rate to be the same as in the distant
past (back billions years).

The present paper is dedicated to search of a possible physical
mechanism leading to the observed variations in the $\beta$-decay
and $\alpha$-decay  rates of radioactive elements and in the
constant of thin structure.

\section{Experiments on observing periodic changes in the $\beta$-decay rate of radioactive elements and constant of thin structure}

In the paper [5], results of measuring the half-period of
$^{32}Si$ ($\beta$-decay) from 1982 till 1986 in the Brookhaven
National Laboratory (USA), are given. In connection with the
uncertainty of this quantity for $^{32}Si$ (60years $\le T_{1/2}
\le$ 700 years), there was used a procedure of concurrent
measurement, by the same detector, of sufficiently stable
$^{36}Cl$ ($T_{1/2} =$ 30800 years [12] with the change in the
decay rate on the level of $10^{-5}$ per four years) reducing to a
minimum systematic errors due to electron equipment drift,
variations in temperature, humidity, asf., and allowing  to
determine $T_{1/2} =$ 172 years for $^{32}Si$.

To compare the results of experiments with those for decays of
other elements, there was chosen the value
$$
U (t) = \left[ \frac{N' (t)}{N'_0 (0)} \right] \exp(+\lambda t),
\, {\text where} \, N' (t) = \frac{dN}{dt} = -\lambda N_0
\exp(\lambda t),
$$

$$
\lambda = \frac{\ln (2)}{T_{1/2}}; \quad t \quad  \mbox {is time}.
$$

In Ref. [5,13], the authors accepted $\lambda = 4.0299 \cdot
10^{-3}$~year$^{-1}$ for $^{32}Si$. According to the exact
exponential law of decay, $U (t) = 1$. Deviations of $U (t)$ from
1 indicate to an inexactness of this law. The results of
measurements for the ratio $^{32}Si / ^{36}Cl =  N' ( ^{32}Si) /
N'(^{36} Cl)$  are shown in Fig. 1 of the work [13]. Each point
corresponds to an average magnitude of the function $U (t)$ over 6
days. As is seen from the Figure, unexpected yearly oscillations
in the $\beta$-decay rate in the system $^{32}Si/^{36}Cl$, were
observed.

In Ref. [6] results of $T_{1/2}$ measurements carried out in
Germany, at the "Physicalisch-Technische Bundesanstalt (PTB)" were
presented, with the use of an ionization chamber for $^{226}Ra$
($\alpha$-decay, $T_{1/2} \approx$ 1602 years) that was used as a
calibration source (a decrease in the average flow of particles on
the level of $0.005\%$ per year) to measure half-periods of decay
for $^{152}Eu$ and $^{154}Eu$ as well as to investigate stability
of semiconductor detectors on the base of super-pure $Ge$ and
$Ge(Li)$. In the same experiments, unexpected yearly oscillations
in the decay rate of $^{226}Ra$ on the level of $(0.10-0.25)\%$
were also recorded. Figure 2 from the work [13] demonstrates $U
(t)$ changes for the flow of $\alpha$-particles in the course of
15-year experiment from 1983 till 1998. Each point in the Figure
corresponds to the average magnitude of the function $U (t)$
through 2.8 days (1968 points in all).

It is seen from the Figs.1-2 from the work [13] that the season
variations observed in the decays of the system $^{32}Si /
^{36}Cl, \, ^{226}Ra$ as well as the changes in the thin structure
constant, are very similar. In Fig. 2
 an angular histogram is
shown for the season distribution of maximum and minimum decay
rates over the Earth orbit in the process of its motion around the
Sun for a 15-year experiment with $^{226}Ra$ (Fig. 2[13])  The
maxima are denoted by arrows, minima  by black rectangles. The Sun
is in the center of histogram. On the line drawn from the portion
of the orbit with the observed decay peaks and valleys to the
place of the Sun, the years of observation for that orbit portion
are given. If the minimum in May, 1990 excluded, all other extrema
in the decay rate are arranged between the lines "mid-June -
mid-December" and "mid September - mid-March". The maximum in the
angular histogram considered corresponds to the line "mid-July -
mid-January".

Contrary to the works [5,6,13] where, in the course of many year
standing experiments on measuring half-decay periods of
radioactive elements, the flows of particles were averaged over
some days to obtain one point, in Refs. [7-10] (Figs. 1,2) the
results of long-term investigating $\beta$-decay rates for
$^{90}Sr$ [8,10], $^{137}Cs$ and $^{60}Co$ [7,9,10], with duration
of measurements from two weeks to five months, were given for
integrated flows of particles over 10 s and one minute exposition.

In these researches, scintillation and $Ge(Li)$ $\gamma$-detectors
for studying $\gamma$-quanta accompanying the $\beta$-decay rate
of radioactive elements, and a fast plastic scintillator YAG:Ce
for direct detecting $\beta$-electrons in the case of $^{90}Sr$
decay investigation, were used. Changes in the $\beta$-decay rate
with periods around 24 hours (the spacings between extremum
deviations could be also 18.5 and 30 hours) and 27 days, were
recorded.  The use of scintillation procedure led to detecting
changes much greater than $5\%$ - level of significance, and the
procedure using $Ge(Li)$ detectors, in a parallel experiments
carried out simultaneously at INR RAS (Troitsk) and JINR (Dubna)
from March 15, 2000 till April 10, 2000, gave for the flow of
$\gamma$-quanta (accompanying, for example, the $\beta$-decay of
$^{60}Co$) deviations more than $1\%$ from their average value.
The deviations in the $\beta$-decay rate changes of radioactive
elements set off three directions in space when rotating the
laboratory system of coordinates together with the Earth. These
three directions are restricted by tangent lines drawn to the
Earth parallels at the place of laboratory position at the time
point when the maximum effect is observed (Figs. 3-7 [9, 10]). As
is seen from Figs. 5,7, the direction corresponding to the
tangents with numbers 8, 9, 19, 20 in Fig. 6 and to numbers 1, 2b,
4, 14, 18, 34, 42, 44 in Fig. 7, is practically coincident with
the direction of the line drawn from the place of the orbit with
the maximum numbers of minimum observations (in July) to the place
of observing the most number of maxima (in January) for the decay
shown in Fig. 2.

\section{Analysis of variations in decays of radioactive elements, thin structure constant, and anisotropy of physical space}

In Ref. [13], an analysis of experiments on investigating periods
of half-decay of radioactive elements was made [5, 6]. It was
shown that the observed yearly oscillations in the decay rate for
the system $^{32}Si/^{36}Cl$ had a coefficient of correlation K
with the changes in the magnitude of $1/R^2$ where $R$ was the
distance between the Earth and the Sun ($K = 0.52$), and the
coefficient of correlation between decay rate of $^{236}Ra$ and
the changes in $1/R^2$ was equal to 0.66. Therewith the
correlation coefficient between decay rate variations for two runs
of observations was 0.88 (Fig. 1-2). In the average the maxima of
decays were observed in mid-January (the perihelion of the Earth
orbit falls on 4-th of January), and the minima were around July
(the apogee is in July, 4).

 To explain the
phenomenon, several hypotheses were proposed [13,14]. In
particular, consideration was given to the hypothesis of global
scalar field that is often used in the physics of unified theories
of interaction between elementary particles as well as in the
theory of expanding Universe [15,16]. In so doing, it is assumed
that the role of a scalar potential is played by the local
gravitational potential of the Sun. One introduces the dependence
$\delta \alpha (r) / \alpha = k_{\alpha} \Delta U (r)$ where
$\Delta U (r)$ is the change in the gravitation potential of the
Sun, and $k_{\alpha}$ is a gain factor associated with the action
of some {\it fifth} force. Its action, as was estimated in many
theories [17-19], is many orders of magnitude less than that of
gravitation which is in turn $10^{38}$ times weaker as compared
with the electromagnetic forces. As is seen from Figures 1-2 from
the work [9] and Fig.7 from [10], the new force revealing itself
in the $\beta$-decay rate changes, should be characterized by a
dimensionless constant non less than $\sim 10^{-14}$ - the
constant  of weak interaction (the latter is dimensional in
standard theories, and the dimension-less value was obtained with
the use of the Compton wave length for the pion [20]).

It should be noted that aside from the works of the type of
[17-19] dedicated to the fifth force, there exist a diversity of
investigations on the new hypothetical anisotropic interaction
[21-33] that has no specific constant but is described by a series
in changes of the modulus of the summary potential $A_{\sum}$ the
magnitude of which cannot exceed the modulus of a new fundamental
constant - the cosmological vector potential $A_g$ first estimated
in the work [34].

The nature of the new force is the follows. In accordance with the
byuon theory [30-33] (a gauge-less theory of formation of physical
space and elementary particles from unobservable objects, the
byuons, the basic characteristic of which is the vector $\vec
A_g$), the potentials of physical fields, in contrast with gauge
theories, are determined values due to violation of gauge
invariance. A part of mass of each elementary particle associated
with the formation of its internal physical space, is proportional
to the modulus of $\vec A_{\sum}$ that cannot be more than the
modulus of $\vec A_g$ equal to $1.9 \cdot 10^{11}$~G cm. As a
consequence, if we decrease in some manner the modulus $|\vec
A_{\sum}|$, each body having mass will be ejected from the region
with the weakened $|\vec A_{\sum}|$ under the action of the new
force. In particular, the modulus $|\vec A_{\sum}|$ can be reduced
by the vector potential of one or another magnetic system.

The anisotropic properties of the new force caused by the
existence of vector $\vec A_g$, were investigated experimentally:

- with the aid of torsion balances arranged in high-current
magnets with the field up to 15 T [21-24, 30-33];

- when measuring heat content of  plasma jet in powerful electric
discharges in accordance with the attitude of discharge current
relative to stars [25-28, 30-33];

- in experiments with gravimeters with attached magnets that did
not change the characteristics of gravimeters [29,30-33];

- when studying a number of terrestrial and astrophysical
phenomena: the anisotropy in the distribution of earthquakes over
the surface of the Globe immobile relative to stars [32-33],
distribution anisotropy of solar flares over the Sun immobile
relative to stars [38-41], anisotropy in distribution of pulsar
velocities in the Galaxy [32, 33, 35, 36], etc.

The most accurate measurements of $\vec A_g$ direction were
carried out in the works [26-27] where it was shown that the new
force ejected substance from the region of weakened modulus
$A_{\sum}$ along the cone with an opening close to $100^{\circ}$
around the vector $\vec A_g$ that had the following coordinates in
the second coordinate system: right ascension $\alpha =
293^{\circ} (19^h 32^m) \pm 10^{\circ}$, declination $\delta =
+36^{\circ} \pm 10^{\circ}$. This direction corresponds to the
line of observation of maximum changes in the decays (Fig. 3, the
line of minima-maxima), i.e. to the line {{\it mid-June -
mid-January} as well as to the directions of tangents to the Earth
parallels with the numbers 8, 9, 19, 20 in Fig. 5 and 1, 2b, 4,
145, 18, 34, 42, 44 in Fig. 7 that are practically coincident with
the direction of lines of the vector potential of the Earth
magnetic field.

To illustrate the anisotropy considered, in Fig. 3 shown is a
diagram of motion directions of pulsars in the picture plane (in
projection onto the celestial sphere) related to the ecliptic
plane.

Let us elucidate the action of the new force on pulsars. Pulsars
are  neutron stars formed in the result of explosion of a
Supernova with the duration of an order of $10^{-3}$~sec. As is
shown in the works [32,33,35,36], the observable velocities of
pulsars and their angular distribution can be explained by the
action of the new force that, if exists, must clearly reveal
itself in the process since the magnetic fields of pulsars may be
as great as $10^{12}$~G, and hence the magnitudes of the vector
potential may come close to the modulus of $\vec A_g$. The
reactive effect causes a pulsar to move oppositely to the
direction of the new force.

As is seen from Fig. 3, the main masses of pulsars move along the
reverse cone of new force action which is in correspondence with
the prediction of the byuon theory and the angular opening of the
season arrangement of $^{226}Ra$ decay rate minima also given in
the Figure.

The modulus may be changed not only due to the vector potential of
the magnetic fields but also under the action of Coulomb potential
[37] and gravitational potential $\varphi_r$ that is less than
zero and hence will always lead, at any mechanism of addition of
potentials, to a decrease of $A_{\sum}$.

The contribution of $\varphi_r$ in the change of $A_{\sum}$ can be
easily obtained from the energy relation

\begin{equation}
e_0 |\Delta A_{\sum}| \sin \gamma = m_e \varphi_0,
\end{equation}

\noindent where $e_0$ is the electric charge of the electron,
$m_e$ is its mass, $|\Delta A_{\sum}| $ -  the change of the
modulus $A_{\sum}$ due to gravitational potential $\varphi_0$ of
the Sun on the Earth orbit, $\sin \gamma$ is a parameter estimated
in Refs. [30-33] ($\sin \gamma \approx 1/k$ where $k = 10^{15}$)
and characterizing anisotropic properties of the physical space.

Describing the expression for the gravitational potential in an explicit form we obtain
\begin{equation}
|\Delta A_{\sum}|  =  \frac{G m_e  M_c}{R^2 e \sin \gamma}
\end{equation}

\noindent where $G$ is the gravitation constant, $M_c = 2 \cdot
10^{33}g$ is mass of the Sun, and $R = 1.5 \cdot 10^{13}$~cm - the
average distance between the Earth and the Sun. Substituting the
numerical values of parameters into Eq. 2 we obtain the magnitude
of $ |\Delta A_{\sum}|$ corresponding to the action of $\varphi_0$
onto the process of formation of electron (and other particles)
internal space on the Earth and being of the order of $10^{10}$~G
cm. This is almost half order of magnitude more than the value of
vector potential due to the solar dipole magnetic field on the
orbit of the Earth and almost two orders of magnitude more than
the vector potential of our planet magnetic field [7, 30-33]. That
is, the action of the new force onto the processes in
consideration in the vicinity of the Earth orbit because of
changes in the potential of the solar gravitation field is
deciding but not the only.

In this connection, because of neglect of influence of
vector-potentials from the Earth's and solar magnetic fields, we
see in Fig.1-2 shifts in the experimental results when observing
maxima and minima in the physical measurements due to basic right
effect of the Sun gravitation potential through the action of the
new force.

The difference $\Delta R$ between the distances of the Earth from
the Sun in perihelion (147.5 millions km) and in aphelion (152.6
millions km) is of the order of 5 millions km, and the
corresponding variations of vectorial potential are

\begin{equation}
\delta |\Delta A_{\sum}|  = \frac{G m_e  M_c}{R^2 e \sin \gamma}
\Delta R = 5.6 \cdot 10^8 \, G   cm
\end{equation}

This value is comparable with the vector potential of the Earth as
well with that of the solar magnetic field. Notice that $\delta
|\Delta A_{\sum}| \propto 1 / R^2$ which corresponds to
observation results, and the abrupt change in the decay rate in
spring, 1988 (Fig. 2) as well as the minimum decay deviations in
spring-summer, 1986, were in proximity to the polarity change of
the Sun`s magnetic dipole (22d  11 years cycle), whereas the
minimum deviations and very considerable changes in the decay rate
registered the end 1998 in an experiment with $^{226}Ra$, preceded
the begin of 23d solar cycle (begin 1999) [38]. The latter decay
deviations were, in all likelihood, taken by the authors of
Ref.[6] as development of instabilities in equipment that may be
correct. Of course, the issue of possible connection between
changes in the decay and the 11-year solar cycle invites further
investigations and evidences.

It is worth noting that the standard model of $\alpha$- decay [1]
does not connect this process with any force, but explains it by
the pure quantum phenomenon, namely the tunnel effect as the
penetration of $\alpha$-particles through the Coulomb barrier.
 The new force of Nature acts if there are $|\Delta A_{\sum}|$  and its gradient.
 This situation is realized probably during  $\alpha$-decay. In this case gradient of $|\Delta
 A_{\sum}|$  is caused by the magnetic moment of the nucleus (for example,$^{241}Am$) . As for  $\beta$ -decay this gradient is connected with the magnetic moment of neutron [9, 30-32]).
The existence of  $|\Delta A_{\sum}|$ in the byuon theory [30-32]
changes the characteristic size of a nucleus and as a consequence
changes the value of the Coulomb barrier. This change causes an
increasing (or decreasing) of a number of particles leaving the
nucleus via the tunnel effect. Hence in an analysis of
$\alpha$-decay we must take into account two effects: changes of
the nucleus size depending on the location of the source at the
Earth orbit during its rotation around the Sun and the acting of
the new force. The main reason of the variations of $\alpha$-
decay rates for $^{226}Ra$ (the even-even nucleus) is the change
of the nucleus sizes due to changes of fundamental scales of the
surrounding World ($\sim10^{-17}$ cm and, $\sim10^{-13}$cm
[30-33]).

Thus the changes observed in the decay rate of radioactive
elements (Figs 1, 2)  are of the same nature and can be caused by
the action of the new anisotropic force and by changes of nuclei
sizes in the case of   $\alpha$- decay..

Let us notice also that the above changes in the decay rate, if a
systematic temporal trend for them will be found, can correct the
time scale in the cosmochronology and in the analysis of
astrophysical phenomena.

In our opinion, the basic problems in the modern cosmology remain
the search for the dark matter and investigating the nature of
interaction leading to formation of dark energy. The solution of
this problem will give us a possibility to understand the essence
of $95\%$ of matter in the observable part of the Universe.

In this connection the search and study of new force (or, may be,
new forces) are of fundamental importance. Our efforts in this
direction will hopefully be useful.

The authors are grateful to the RAS academician S. T. Belyaev and
B. V. Komberg for the exchange of information and discussion of
questions of possible relation between the observed changes in the
decay rates of radioactive elements and the anisotropy of physical
space.

\begin{figure}
\includegraphics{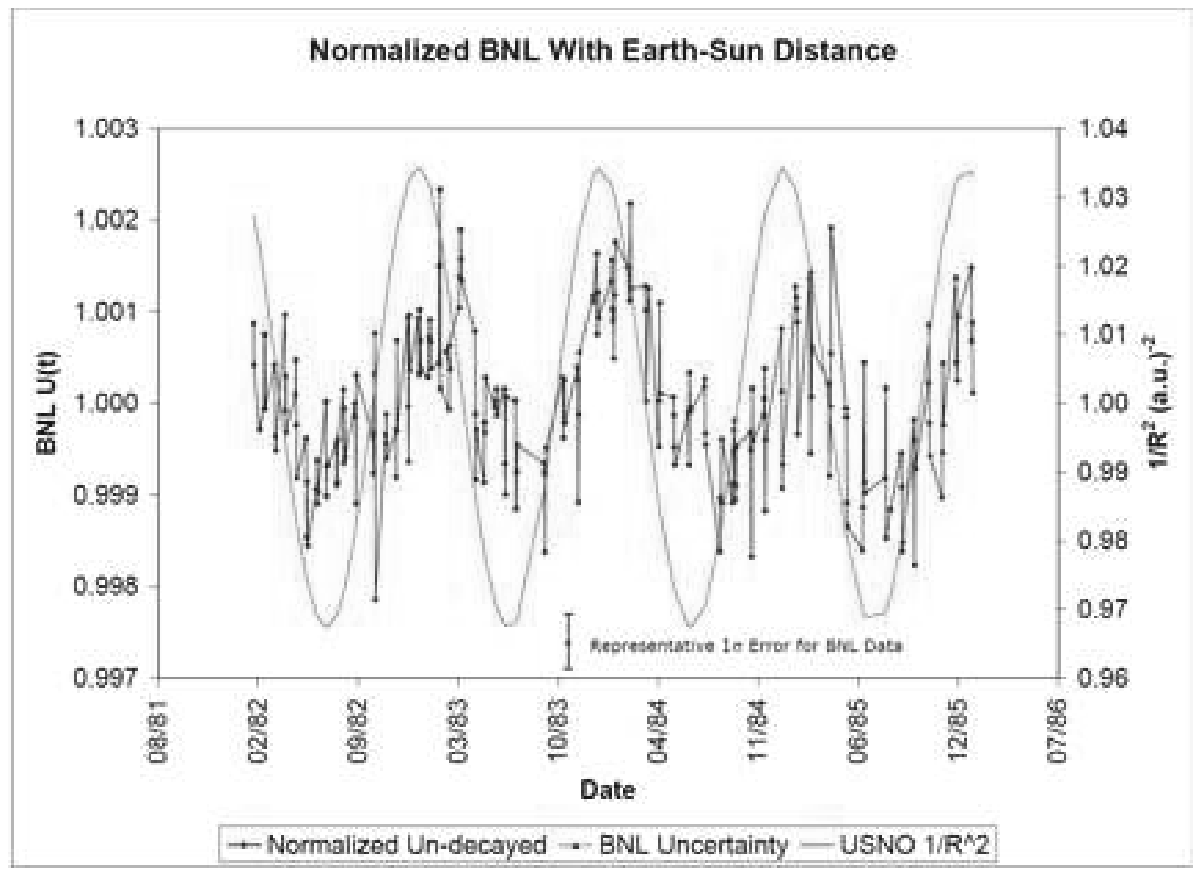} \caption{Unprocessed BNL data
for the time function U(t) plotted for relation $^{32}Si/^{36}Cl$
for the period from August, 1981 (08/81) till July, 1986 (07/86).
The right ordinate is the value $1/R^2$, where R is the distance
from the Earth to the Sun in astronomical units (a.u.). Its
variation with time is shown as a solid line (taken from [13]).
Points are numbers of successive magnitudes of U(t) for the time
point considered; I is experimental error shown at the bottom of
Figure.}
\end{figure}

\begin{figure}
\centerline{\includegraphics{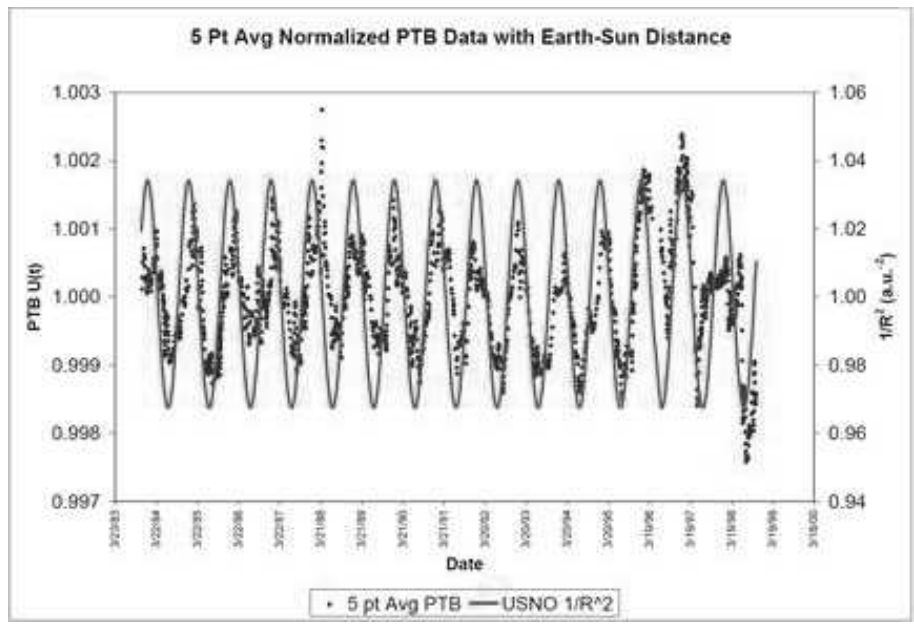}} \caption{The U(t)
data from the RTV laboratory for $^{226}Ra$ depending on time
(abscissa: from March 21, 1983, till March 19, 2000). The right
ordinate is the value $1/R^2$ with R the distance from the Earth
to the Sun in astronomical units (a.u.). Its variation with time
is shown as a solid line (taken from [13]) Points are sliding
averages over 5 data points from the initial series. There are
altogether 1968 successive magnitudes of U(t) for the time period
indicated.}
\end{figure}

\begin{figure}
\centerline{\includegraphics{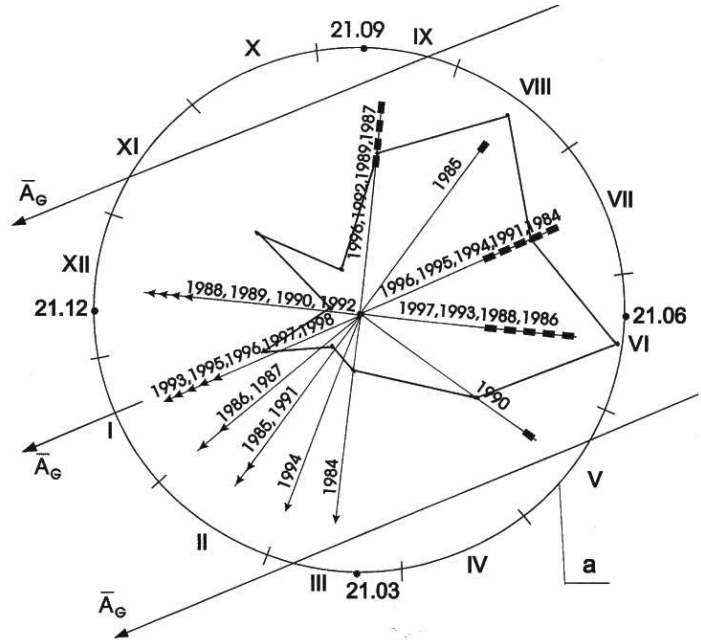}} \caption{ The
angular histogram of season distribution (fourth and further
months of year) of maxima and minima of decay rates on the Earth
orbit (a) in the process of its motion around the Sun for a 15
years experiment with $^{226}Ra$ (Fig. 2). Maxima are indicated by
arrows, minima  by black rectangles. The Sun is in the center of
the diagram. On the line from the orbit portion with decay maxima
or minima to the place of the Sun, the corresponding years are
indicated. The solid line shows angular distribution for velocity
directions of pulsars motion.
      $A_G$ is the cosmological vector potential.
      21.03 etc. are characteristic points of the Earth orbit.}
\end{figure}

\begin{figure}
\centerline{\includegraphics{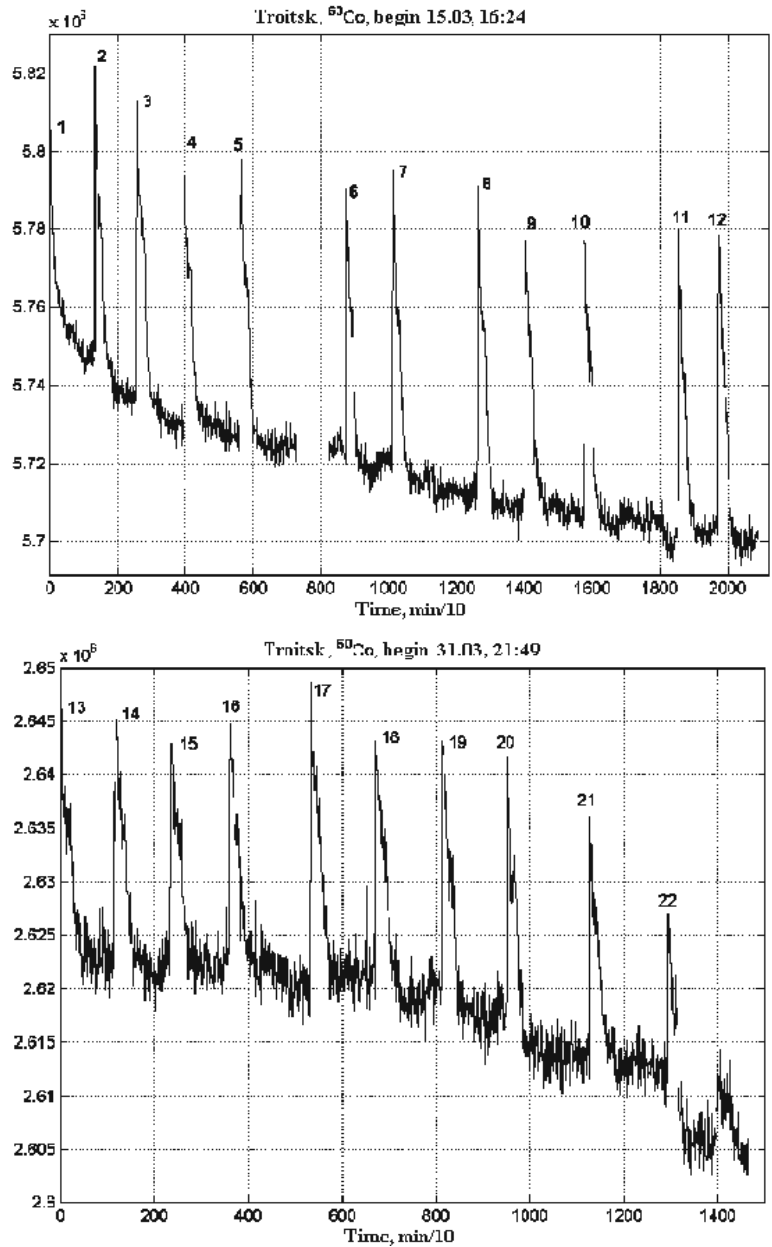}} \caption{ Changes
of $\gamma-quanta$ flux for $\beta-decay$ of   $^{60}Co$ (INR,
Troitsk).}
\end{figure}

\begin{figure}
\centerline{\includegraphics{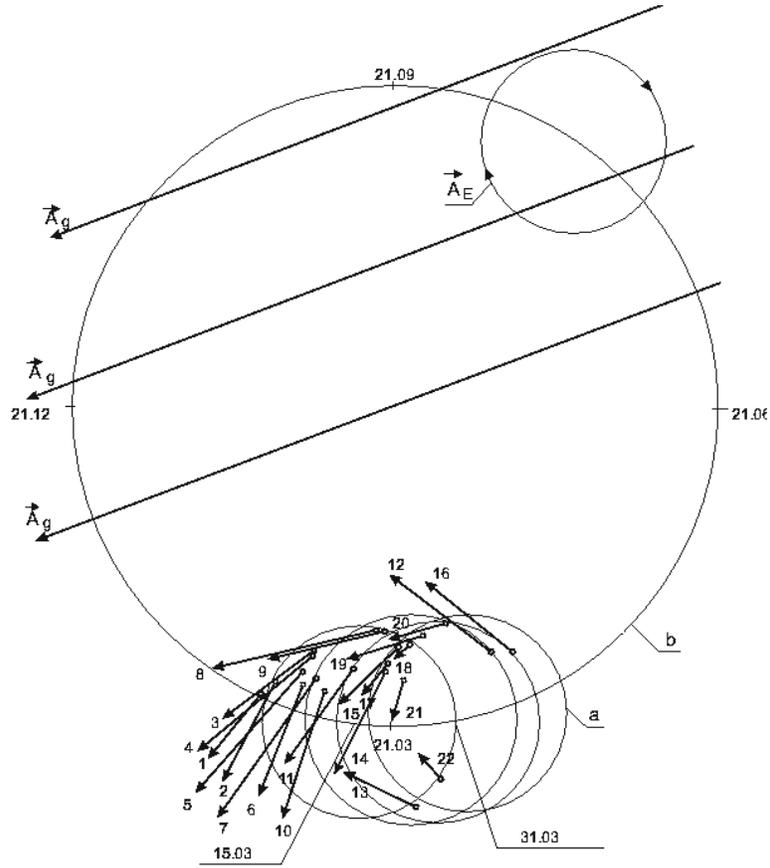}} \caption{Locations
of extrema in  $\gamma$-quanta fluxes for $\beta$-decay of
$^{60}Co$ (INR, Troitsk) (see Fig.4). Arrows are  location of
maximal $\gamma-quanta$ flux with the pointing of the direction of
the new force drawing along the tangent to the Earth  parallel; a
is trajectory of the radioactive source during its rotation in
common with the Earth; b is trajectory of the Earth and the
radioactive source around the Sun; 21.03 etc. are vernal
equinoctial point and other characteristic points of the Earth
trajectory; $A_E$ is direction of the vector potential of the
dipolar Earth magnetic field;
 $A_g$ is direction of the cosmological vector
potential.}
\end{figure}

\begin{figure}
\centerline{\includegraphics{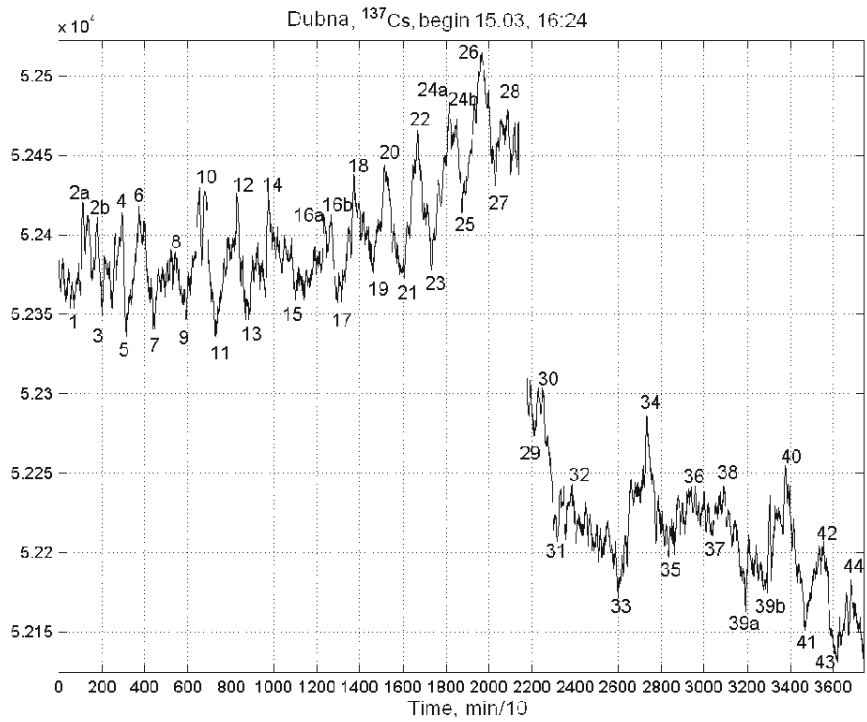}} \caption{ Changes
of $\gamma$-quanta flux for $\beta$-decay of $^{137}Cs$ (JINR,
Dubna). The experiment has been carried out from $16^h 24^m$
15.03.2000 till the midnight of  10.04.2000. Break near the 2200th
point  has been caused by the technical reason connected with the
pouring some more liquid nitrogen. The growing of the curve from
the 1000th till  the 2000th  points is the consequence of the
small amount of nitrogen in the detector.}
\end{figure}

\begin{figure}
\centerline{\includegraphics{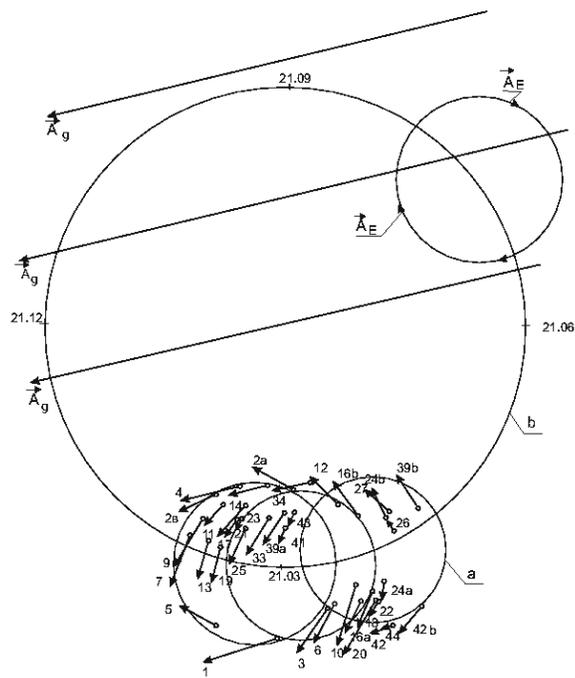}} \caption{The same
as Fig. 6 but for $\gamma$-quanta flux from $^{137}Cs$
 minima and
maxima (JINR, Dubna)}
\end{figure}

\end{document}